\documentclass[sigconf]{acmart}
\usepackage{algorithm}
\usepackage{graphicx}
\usepackage{multirow}
\usepackage{booktabs}
\usepackage[tight,footnotesize]{subfigure}
\usepackage{algorithmic}
\usepackage{amsfonts}
\usepackage{amsmath}

\AtBeginDocument{%
  \providecommand\BibTeX{{%
    \normalfont B\kern-0.5em{\scshape i\kern-0.25em b}\kern-0.8em\TeX}}}

\setcopyright{acmlicensed}
\copyrightyear{2024}
\acmYear{2024}
\acmDOI{XXXXXXX.XXXXXXX}

\begin{document}

\title{Hybrid Heterogeneous Clusters Can Lower the Energy Consumption of LLM Inference Workloads}

\author{Grant Wilkins}
\email{gfw27@cam.ac.uk}
\affiliation{%
  \institution{University of Cambridge}
  \city{Cambridge}
  \country{UK}
}


\author{Srinivasan Keshav}
\email{sk818@cam.ac.uk}
\affiliation{%
  \institution{University of Cambridge}
  \city{Cambridge}
  \country{UK}
}

\author{Richard Mortier}
\email{rmm1002@cam.ac.uk}
\affiliation{%
  \institution{University of Cambridge}
  \city{Cambridge}
  \country{UK}
}

\begin{abstract}
Both the training and use of Large Language Models (LLMs) require large amounts of energy. Their
increasing popularity, therefore, raises critical concerns regarding the energy efficiency and sustainability of data centers that host them. This paper addresses the challenge of 
reducing energy consumption in data centers running LLMs. We propose a \textit{hybrid} data center model that uses a cost-based scheduling framework to dynamically allocate LLM tasks across hardware accelerators that differ in their energy efficiencies and computational capabilities. 
Specifically, our workload-aware strategy determines whether tasks are processed on energy-efficient processors or high-performance GPUs based on the number of input and output tokens in a query. Our analysis of a representative LLM dataset, finds that this hybrid strategy can reduce CPU+GPU energy consumption by 7.5\% compared to a workload-unaware baseline.

\end{abstract}


\keywords{Large-Scale Data, LLMs, Energy Consumption, Green Computing}

\maketitle

\section{Introduction}

Large Language Models (LLMs) such as OpenAI's GPT-4~\cite{openai2023gpt4} and Google's PaLM~\cite{anil2023palm} have become emblematic of the AI revolution, driving significant advancements not only in natural language understanding, generation, and translation but also in summarizing and contextualizing large volumes of textual data. Characterized by their extensive scale and depth, their deployment demands substantial computational resources and hence poses significant challenges in terms of energy consumption and operational efficiency~\cite{wu2022sustainable}. The increasing application of LLMs across diverse sectors further compounds these challenges, because datacenters, which are responsible for a considerable portion of global electricity consumption, must balance performance targets for LLM tasks running on heterogeneous hardware with the need for energy efficiency~\cite{luccioni2022estimating, chien2023reducing}. Reducing the energy efficiency of LLMs thus emerges as both a technical challenge and an environmental imperative~\cite{MYTTON20222032}. 

Traditional data center designs often struggle to best exploit the capabilities of heterogeneous hardware-based LLMs, particularly when trying to minimize energy consumption without sacrificing output quality and latency~\cite{chen2024landscape}. However, this challenge also presents an opportunity to innovate in datacenter architecture and management. We show that by rethinking how GPU resources are allocated and managed, there is potential to significantly reduce the energy footprint of LLM deployments while maintaining or even enhancing computational performance.

We find that a dynamic task-scheduling model that assigns LLM tasks to GPUs based on the resulting energy efficiency can reduce overall energy. Moreover, implementing a workload-aware system for input and output token processing can further reduce energy usage.
Thus, a hybrid datacenter task allocation model, 
which allocates different tasks to different hardware accelerators based on their system demands, can reduce the overall energy consumption of LLM inference compared to a workload-unaware baseline.

Our contributions are as follows:
\begin{enumerate}
    \item We analyze the energy consumption and runtime of several 7B-parameter LLMs' across various hardware configurations.
    \item We propose and evaluate a workload-aware scheduler for LLMs that optimizes energy efficiency based on the size of input and output token loads, demonstrating a 7.5\% decrease in energy consumption over non-workload-aware baselines.
    \item We release a comprehensive dataset and benchmark suite for evaluating the energy efficiency of LLM inference, enabling researchers and practitioners to assess the impact of their design choices.
\end{enumerate}

Through these contributions, we hope to support more sustainable and cost-effective AI inference deployments.

The remainder of this paper is as follows: Section~\ref{sec:background} provides background information on LLM inference and energy consumption in AI systems. Section~\ref{sec:problem} formulates the problem and introduces our cost function. Section~\ref{sec:methods} details the methods used for benchmarking LLM inference on diverse systems. Section~\ref{sec:results} presents the performance results of LLM inference across multiple hardware configurations. Section~\ref{sec:analysis} proposes and evaluates our energy-optimal hybrid data center design. Finally, Section~\ref{sec:related-works} discusses related works, and Section~\ref{sec:conclusions} summarizes the conclusions of the paper.

\section{Background}\label{sec:background}

\subsection{Inference Using Large Language Models}
Transformer-based neural network architectures have led to impressive gains
in the performance of LLMs for language understanding and generation~\cite{bommasani2022opportunities}. 
LLMs such as OpenAI's GPT-4~\cite{openai2023gpt4} and Google's Gemini~\cite{geminiteam2024gemini} have demonstrated 
human-level proficiency on many language benchmarks while requiring billions of parameters and massive datasets for training. The inference phase of LLMs involves utilizing a trained model to make predictions based on new, unseen data. 
Unlike the training phase, which is typically a one-time, compute-intensive process that occurs offline, 
inference is an ongoing, real-time process that directly impacts end-user experiences~\cite{chien2023reducing}. 
This phase is critical as it represents the point at which AI capabilities become accessible to users.

Inference in LLMs can be computationally expensive due to several factors: \textbf{(1) Model Size:} The sheer size of these models, often billions of parameters, necessitates significant computational power to process each query~\cite{wu2022sustainable}. 
\textbf{(2) Latency Expectations:} Many applications based on LLMs, such as digital assistants, automated writing aids, and real-time translators, require low-latency responses~\cite{wang2024efficient}. 
\textbf{(3) Scalability:} The ability to scale inference operations to accommodate varying user demands without degradation in response times is crucial. 

\subsection{Energy Consumption in AI Systems}
Recent reports have found that the computational requirements for state-of-the-art AI entail massive energy consumption and carbon emissions~\cite{samsi2023words, chien2023reducing, wu2022sustainable, luccioni2022estimating, patterson2021carbon}. 
The energy intensity of AI systems can be broadly divided into the energy required for training versus inference after models are deployed~\cite{henderson2020towards}. Training complex models on massive datasets is an energy-intensive process, with estimates finding that training GPT-3 required 1,287 megawatt-hours of energy~\cite{patterson2021carbon}. 
LLMs can  also have huge emissions depending on deployment scale and hardware efficiency~\cite{samsi2023words}. 
For example, over a year of use, inference by LLMs on cloud infrastructure can consume over 25$\times$ more energy than training a model~\cite{chien2023reducing}. Optimizing software and hardware specifically for AI workloads is thus essential~\cite{anderson2023treehouse}.

\subsection{Heterogeneous Systems for Efficient Computing}
Modern systems demonstrate a complex interplay between scale, architecture, workload behavior and efficiency objectives. 
The architecture of compute nodes can significantly impact the energy efficiency and processing capabilities of large-scale computing systems~\cite{lang2010wimpy}. Conventional server architectures based on multicore CPUs face energy proportionality and scalability limitations for modern data-intensive workloads~\cite{lo2015heracles}. 
Several researchers have explored heterogeneous server configurations to improve energy efficiency~\cite{hu2021minimizing,he2011position,hussain2021energy, liu2023online}. 
Distributed solutions can translate to lower energy efficiency, as communication overheads dominate~\cite{georgiou2015adaptive}. 
Still, specialized clusters like NVIDIA's DGX show 4x better performance per watt over conventional servers~\cite{spetko2021dgx}. 

\section{Problem Formulation}\label{sec:problem}

To model the operational demands of a hybrid, heterogeneous datacenter hosting LLMs, 
we define a cost function to reflect the workload distribution across different systems. 
We define a cost function \( U(m, n, s) \) that accounts for both energy consumption and runtime:
\begin{equation}
    U(m, n, s) = \lambda E(m, n, s) + (1-\lambda) R(m, n, s),
\end{equation}
where \( m \) and \( n \) denote the number of input and output tokens, respectively. \( \lambda \in [0,1] \) is a tunable parameter that balances the weight of energy efficiency versus speed. \( E(m, n, s) \) is the energy consumed by system \( s \) to process \( m \) input tokens and generate \( n \) output tokens, measured in joules. \( R(m, n, s) \) is the time required to process these tokens on system \( s \), measured in seconds.

Our objective is to minimize the total cost across all tasks and systems:

\begin{align}\label{eqn:optimization}
     \min_{\{Q_s\}_{s \in S}} &\sum_{s \in S} \sum_{(m, n) \in Q_s} U(m, n, s) \\ &\text{s.t.} \quad \bigcup_{s \in S} Q_s = Q \\  &\forall s: Q_s \cap Q_{s'} = \emptyset \text{ for } s \neq s'
\end{align}
where \( S \) is the set of all systems, \( Q \) is the total set of queries, \( Q_s \) is the subset of queries assigned to system \( s \).

This model ensures that each query is processed exactly once, optimizing for energy efficiency or quick response times, depending on the operational needs, as parameterized by \( \lambda \). 
We note, however, that certain systems may be better suited to specific tasks, based on the workload characteristics, such as the need for rapid response times. Adjustments in \( \lambda \) allow the datacenter to shift its focus between minimizing energy consumption and reducing runtime as operational priorities change.

\section{Methods}\label{sec:methods}
Here, we describe the methods and tools we use to benchmark LLM inference. In all cases, we use Huggingface's Accelerate~\cite{accelerate} to standardize hardware optimization for inference across all platforms. T
his library takes advantage of the available accelerator resources and shards models accordingly to minimize intermediate communication and maximize the distributed capabilities for computation across the devices.




\subsection{Model Selection}\label{subsec:models}

Our study employs three open-source LLMs for their capabilities and ability to run on diverse hardware efficiently: (1) Falcon~\cite{falcon}, (2) Llama-2~\cite{llama2}, and (3) Mistral (7B parameters)~\cite{Mistral}. These models were selected to represent a spectrum of architectures and training corpora. We subject each model to a series of standardized NLP tasks to evaluate their energy consumption during inference. 

\subsubsection{Falcon}
The Falcon (7B)~\cite{falcon} model utilizes multi-query attention, significantly reducing memory requirements and increasing processing speed. The model's training on the bilingual RefinedWeb dataset enhances its applicability across diverse linguistic contexts.

\subsubsection{Llama-2}
We select Llama-2 (7B) for its optimization in dialogue tasks and its improvements in safety and helpfulness. The model's unique pretraining methodologies and advanced architectural features, such as grouped-query attention, make it an ideal candidate for analyzing energy efficiency in complex language tasks.

\subsubsection{Mistral}
 We include Mistral (7B)~\cite{Mistral} for its grouped-query attention and sliding window attention mechanisms, contributing to fast and efficient inference. Its superior performance in various benchmarks, especially in reasoning, mathematics, and code generation, makes it an essential model for our analysis.

\subsection{Energy Profiling of Diverse Systems}

Depending on the platform, we profile each system's energy consumption during inference using customized setups that capture runtime and energy or power metrics. Here, we describe how we monitor the energy usage of NVIDIA GPUs, Apple Silicon CPU/GPU, Intel CPUs, and AMD CPUs.

\subsubsection{NVIDIA GPUs}
We use PyJoules~\cite{pyJoules}, a Python-based energy measurement library, to quantify the energy consumption associated with inference on NVIDIA GPUs. PyJoules provides an interface to \texttt{NVML}~\cite{nvidia_nvml}, providing a software-defined energy usage assessment for targeted NVIDIA devices. This tool offers real-time energy consumption of GPUs for a given tracked process, which is a critical component of our analysis given the GPU-heavy computation involved in LLM inference.

\subsubsection{Apple Silicon CPU/GPU}\label{sec:apple-profile}
No standard energy measurement tools are available for profiling energy and power usage for Apple Silicon through an API like PyJoules or RAPL. Therefore, we employ a daemon-based approach to poll macOS' \texttt{powermetrics} utility, providing a detailed view of the energy usage during model inference. To capture the energy consumption of the M1 GPU, we execute the \texttt{powermetrics} command through a Python subprocess. This command returns the percentage of the CPU power each CPU \texttt{top} process uses and the total CPU and GPU power consumption in 200ms intervals. This interval was chosen after testing to find the finest granularity measurement without incurring a significant CPU overhead for the I/O of buffering the large \texttt{powermetrics} output into memory. 

The energy monitoring is conducted concurrently with the LLM inference. A separate thread is dedicated to running the \texttt{powermetrics} command, ensuring real-time data collection. Post-inference, the collected data is processed to extract the recorded power data and then find the energy consumption through integration over the runtime. The GPU energy consumption, $E_{Total, GPU},$ is straightforward to calculate for each recorded power value, $P_{GPU, i},$ at each timestep $\Delta t_i.$
\begin{equation}
    E_{Total, GPU} = \sum_{i}{P_{GPU, i} \Delta t_i}.
\end{equation}
The CPU power draw data is less clear, as many processes run on the CPU. However, an "energy impact factor" through \texttt{powermetrics} allows us to infer how much power our Python inference process uses. Therefore, we calculate the CPU energy,  $E_{Total, CPU},$ by multiplying $P_{CPU, i}$ by the "energy impact factor," which we denote as $\alpha_i,$ at each timestep:
\begin{equation}
    E_{Total, CPU} = \sum_{i}{ ( \alpha_i P_{CPU, i}) \Delta t_i}.
\end{equation}

\subsubsection{Intel CPUs}
For Intel CPUs, we leverage PyJoules, a Python-based energy measurement library similar to our approach for NVIDIA GPUs. This tool supports RAPL (Running Average Power Limit) interfaces, enabling us to obtain fine-grained energy consumption data~\cite{rapl}. We focus on two primary RAPL domains: Package 0 and Package 1, which correspond to the entire CPU package's energy consumption, including all cores in the package.

PyJoules allows us to capture the energy usage of these domains in real time, enabling us to profile the energy consumption specifically during model inference tasks. To account for base energy consumption unrelated to our inference process, we conduct a pre-analysis phase to measure the CPU's average idle power draw. This idle measurement is then subtracted from the total energy consumption during inference to accurately determine the net energy expenditure attributable to the inference process.

We instrument our code to query the RAPL readings at the start and end of the inference task, calculating the energy consumption as follows:
\begin{equation}
    \begin{split}
E_{Total, CPU} = \sum_{i}{\bigg(}&\left(P_{Package-0, i} - P_{Package-0, Idle}\right) \\
&+ \left(P_{Package-1, i} - P_{Package-1, Idle}\right)\bigg)\Delta t_i,
\end{split}
\end{equation}
where $P_{Package-0,i}$ and $P_{Package-1,i},$ represent the power draw from Package 0 and Package 1, respectively, and $P_{Package-0,Idle}$ and $P_{Package-1,Idle}$ represent the average idle power draw of the CPU packages, respectively.

\subsubsection{AMD CPUs}
We adopt a different strategy for AMD CPUs due to the absence of a Python API. Instead, we utilize AMD$\mu$Prof's \texttt{timechart} feature, which provides detailed power draw metrics for every core on the chip at fine-grained intervals. By polling AMD$\mu$Prof at 100ms intervals, we can capture the power draw of each physical core throughout the model inference process.

To ensure we accurately attribute the energy consumption to our inference task, we monitor the CPU core residency through \texttt{psutil}. This information allows us to identify and record the specific cores actively engaged in the inference process at each time step. The total energy consumption for the inference task is then calculated by summing the power usage across all active cores and summing over the product of the power usage and time of inference, as follows:
\begin{equation}
E_{Total, CPU} = \sum_{core}{\left(\sum_{i}{P_{core, i} \Delta t_i}\right) }
\end{equation}
where $P_{core, i}$ represents the power draw of an individual core at each time step, $i$.

\section{LLM Inference Performance on Diverse Clusters}\label{sec:results}
\subsection{Hardware and Software Versions}
The systems we profile are shown in Table~\ref{tab:systems}. We consider these systems as they demonstrate three prominent CPU manufactures and different generations of GPUs. We utilize PyTorch v2.0.1, Torchvision v0.15.2, Numpy v1.26.0, Huggingface v0.20.2, and Accelerate v0.26.1.

\begin{table*}[!htb]
\centering
\begin{tabular}{@{}lcccc@{}}

\toprule
System Name & CPU & GPU(s) per Node & DRAM per Node& VRAM per GPU\\ \midrule
Macbook Pro & 10-core M1 Pro & 14-core M1 Pro & 32GB & - \\
Swing AMD+A100  & 2$\times$64-core AMD EPYC 7742 & 8$\times$NVIDIA A100 & 1TB & 40GB\\
Palmetto Intel+V100 & 40-Core Intel Xeon 6148G & 2$\times$NVIDIA V100 & 376GB & 16GB\\
\bottomrule
\end{tabular}
\caption{Our System Configurations}
\label{tab:systems}
\end{table*}

We note that the M1-Pro results only include the Llama-2 (7B) and Mistral (7B) results, as Falcon (7B) generally did not complete tasks in less than two orders of magnitude greater runtime.

\subsection{Experimental Strategy}

To comprehensively evaluate the performance of different system configurations across various models, we conducted a series of controlled experiments. We systematically varied the number of input and output tokens to measure their effects on runtime and energy consumption under two main experimental conditions. In each experiment we do not allow for key-value caches to be re-used to ensure our testing environment is standardized.

\subsubsection{Vary Input Tokens} For the first experimental condition, we executed inference requests with increasing input token sizes, ranging from 8 to 2048 tokens, while maintaining a fixed output token size of 32. This setup allowed us to isolate the impact of input size on the system's performance and energy efficiency.

\subsubsection{Vary Output Tokens} In the second set of experiments, we varied the output token limit from 8 to 4096 tokens, keeping the input token size constant at 32. This approach helped us understand how increasing output demands affect the runtime and energy consumption of the systems tested.

\subsubsection{Randomization and Stopping Criteria}
Each experiment was conducted in a randomized order to mitigate any potential bias introduced by the sequence of tests. To ensure the reliability of our results, we adhered to strict criteria for statistical confidence. Each configuration was tested repeatedly until either of two conditions was met:
(1) The measured runtime had to be within 0.5 seconds of the actual mean runtime with 95\% confidence. (2) A maximum of 25 trials were conducted for each setting if the first condition could not be met.

\subsection{Input Token Analysis}
Here, we present the impacts on runtime, energy consumption per token, and throughput for LLMs across different hardware configurations while varying the number of input tokens. We perform these experiments using the suite of systems outlined in Table~\ref{tab:systems} with the models outlined in Section~\ref{subsec:models}. In our experiments on the Palmetto Intel+V100 system, the V100 GPU had an out-of-memory error beyond 1024 output tokens for Falcon (7B).

\begin{figure*}[!htb]
  \centering  
  \captionsetup{justification=centering}
\subfigure[{Runtime\hspace{-0.8cm}}]
{
\label{fig:input-runtime}
\raisebox{-1cm}{\includegraphics[scale=0.31]{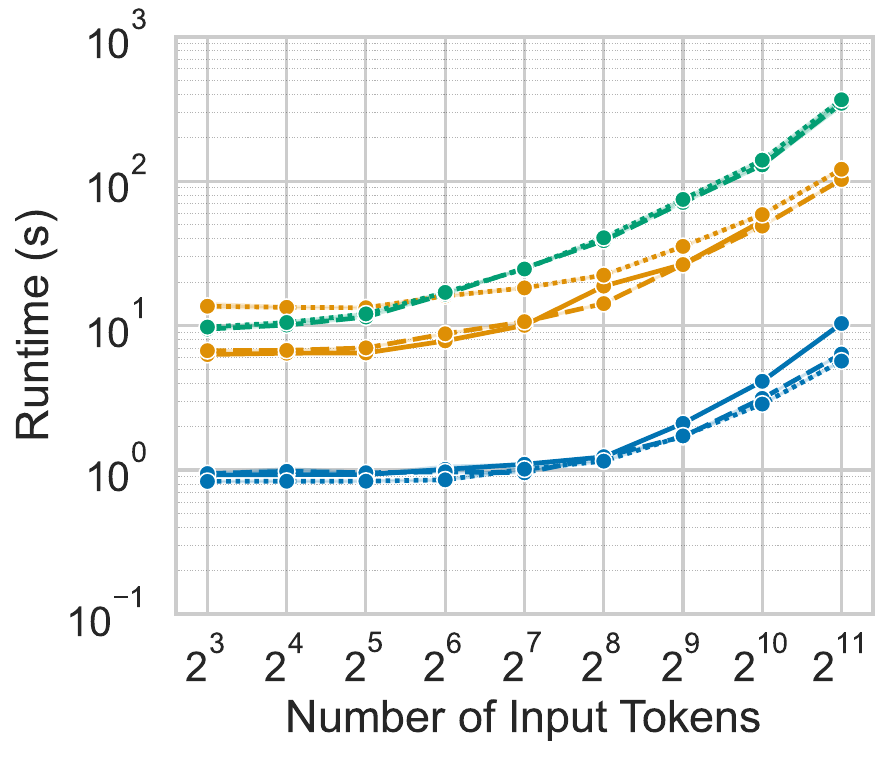}}}
\subfigure[{Throughput\hspace{-0.5cm}}]
{
\label{fig:input-throughput}
\raisebox{-1cm}{\includegraphics[scale=0.31]{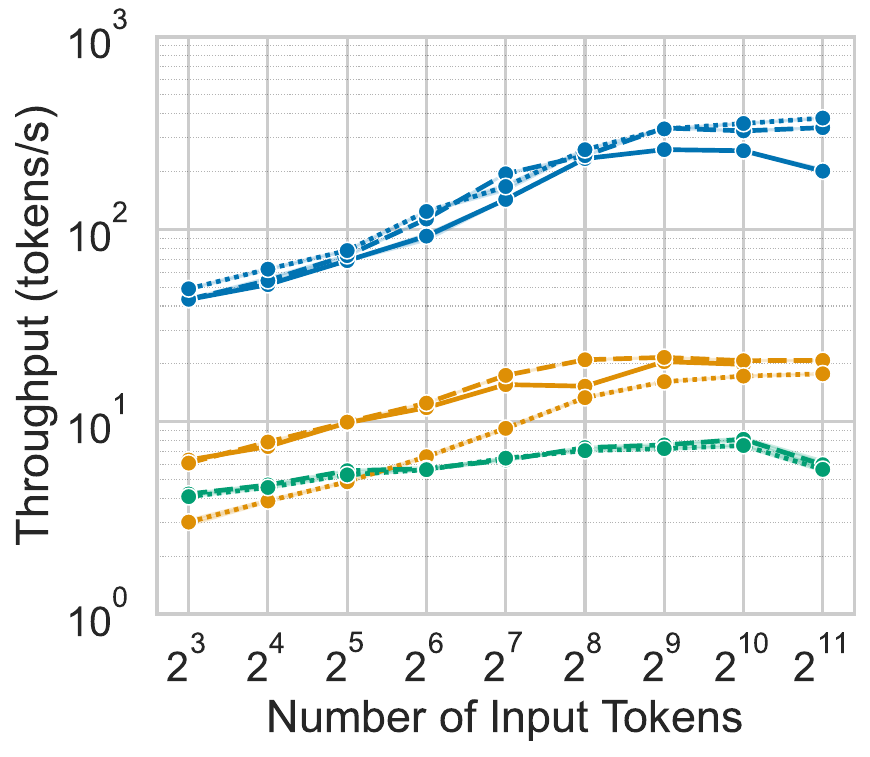}}
}  
\subfigure[{Energy per Token\hspace{2.2cm}}]
{
\label{fig:input-energy-per-token}
\raisebox{-1cm}{\includegraphics[scale=0.31]{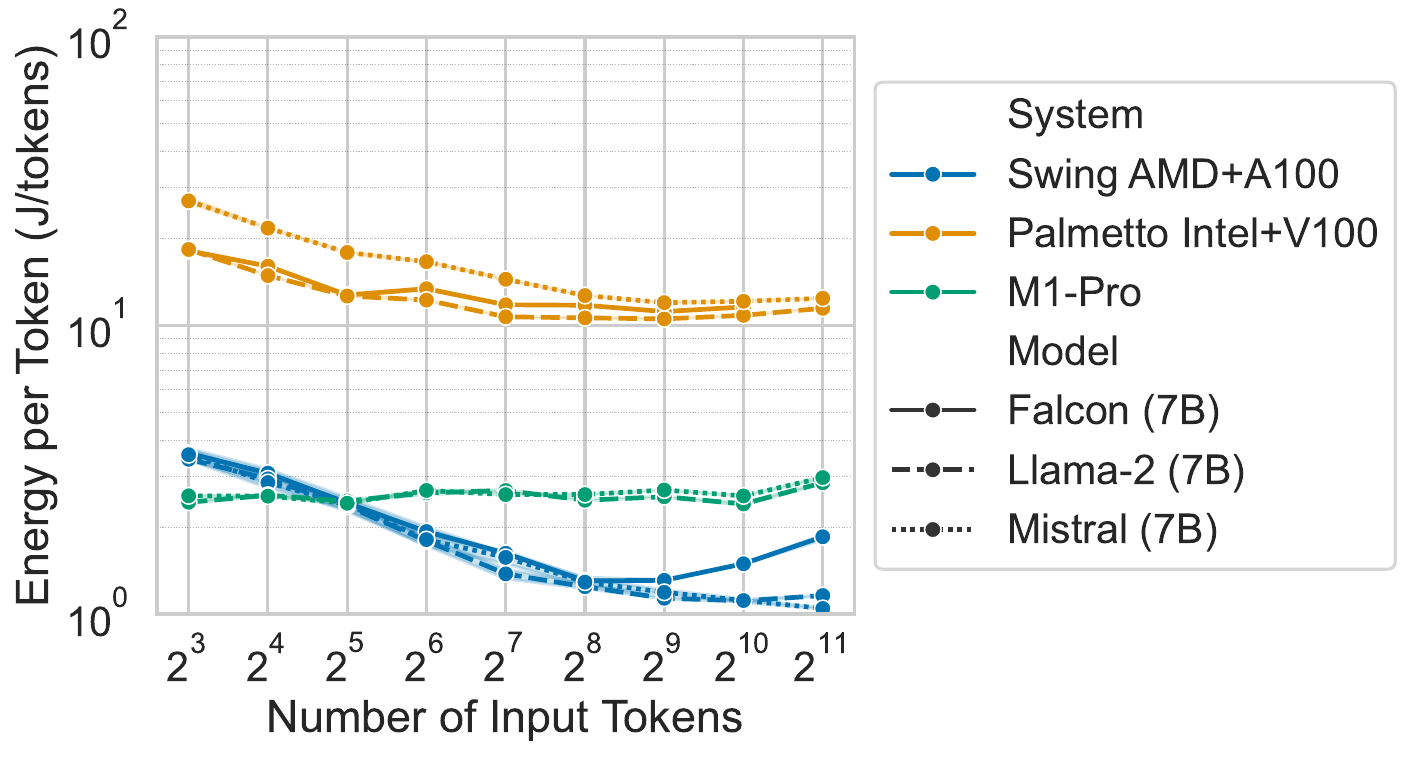}}}
\caption{Performance of Various Systems and Models for Processing Variable Input Tokens--Due to the low variance in the data, error bars are too small to be visible.}
  \label{fig:input-token-results}
  \vspace{-3mm}
\end{figure*}

Our runtime measurements show a significant increase as input tokens grow. As depicted in Figure~\ref{fig:input-runtime}, all systems exhibit a nonlinear escalation in runtime with increasing token counts, with the M1-Pro system showing the most significant magnitude. This trend highlights the computational burden imposed by larger input sizes, particularly on smaller systems that are not as well designed to handle extensive workloads.

For all systems, we notice that throughput follows a "roofline model" with increasing input tokens~\cite{roofline}. Figure~\ref{fig:input-throughput} illustrates these dynamics, indicating an increase in throughput for all systems until a certain point where inference becomes bound by compute and not by the overhead of the software, as described by roofline performance models~\cite{roofline}.

Energy efficiency varies markedly across different systems. The M1-Pro demonstrates consistently low energy consumption per token, particularly for smaller input sizes, as shown in Figure~\ref{fig:input-energy-per-token}. This efficiency reflects the M1-Pro's design optimization for low-power operations. In contrast, the Swing AMD+A100, while capable of handling more significant token inputs more efficiently, consumed more energy per token for small workloads yet became more energy efficient at larger input token sizes, underscoring a trade-off between workload size and energy efficiency.

\subsection{Output Token Analysis}
Here we examine the performance trends associated with increasing the number of output tokens for our LLMs and systems of interest, specifically focusing on runtime, energy consumption per token, and throughput. In our experiments, the M1-Pro also could not generate more than 512 output tokens without significant runtime penalties. For the Palmetto Intel+V100 system, the V100 GPU had an OOM error beyond 1024 output tokens for Falcon (7B) and for all models beyond 2048 tokens.

\begin{figure*}[!htb]
  \centering  
  \captionsetup{justification=centering}
\subfigure[{Runtime\hspace{-0.8cm}}]
{
\label{fig:output-runtime}
\raisebox{-1cm}{\includegraphics[scale=0.31]{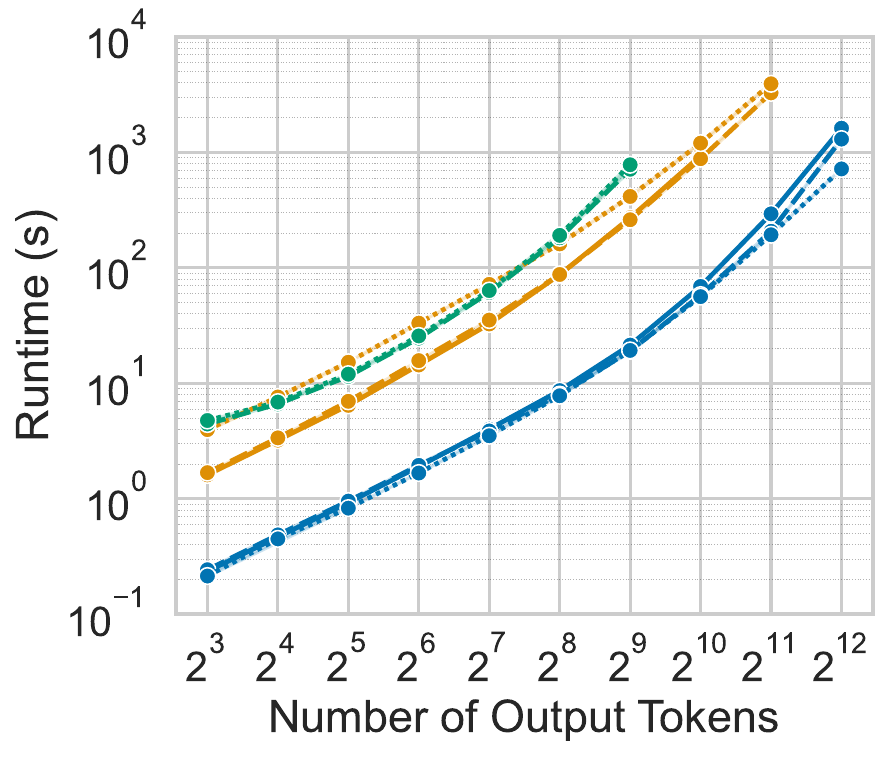}}}
\subfigure[{Throughput\hspace{-0.5cm}}]
{
\label{fig:output-throughput}
\raisebox{-1cm}{\includegraphics[scale=0.31]{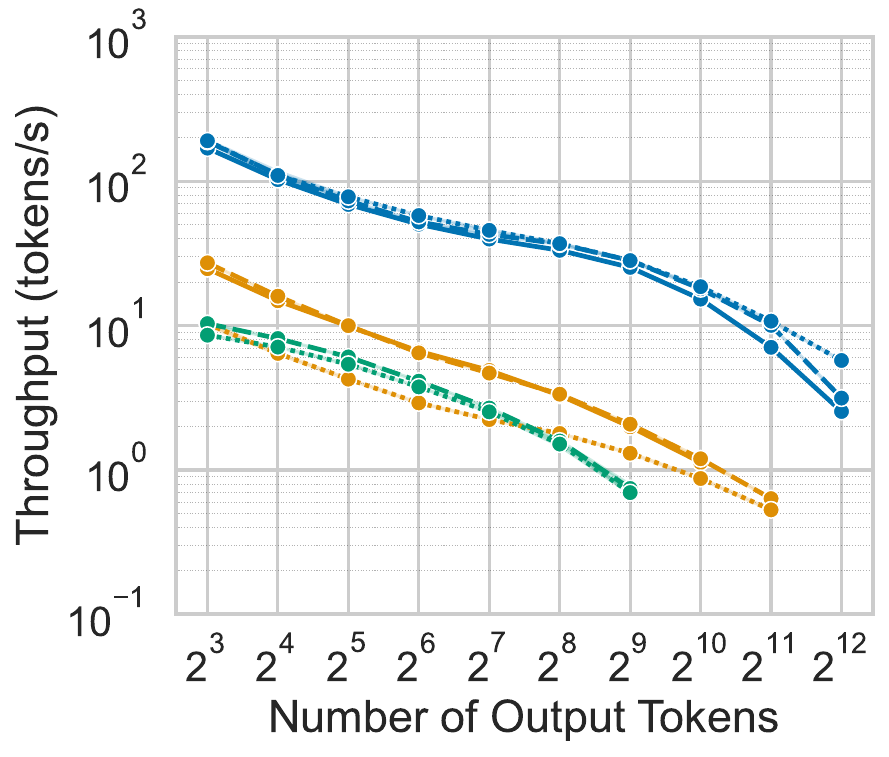}}
}  
\subfigure[{Energy per Token\hspace{2.2cm}}]
{
\label{fig:output-energy-per-token}
\raisebox{-1cm}{\includegraphics[scale=0.31]{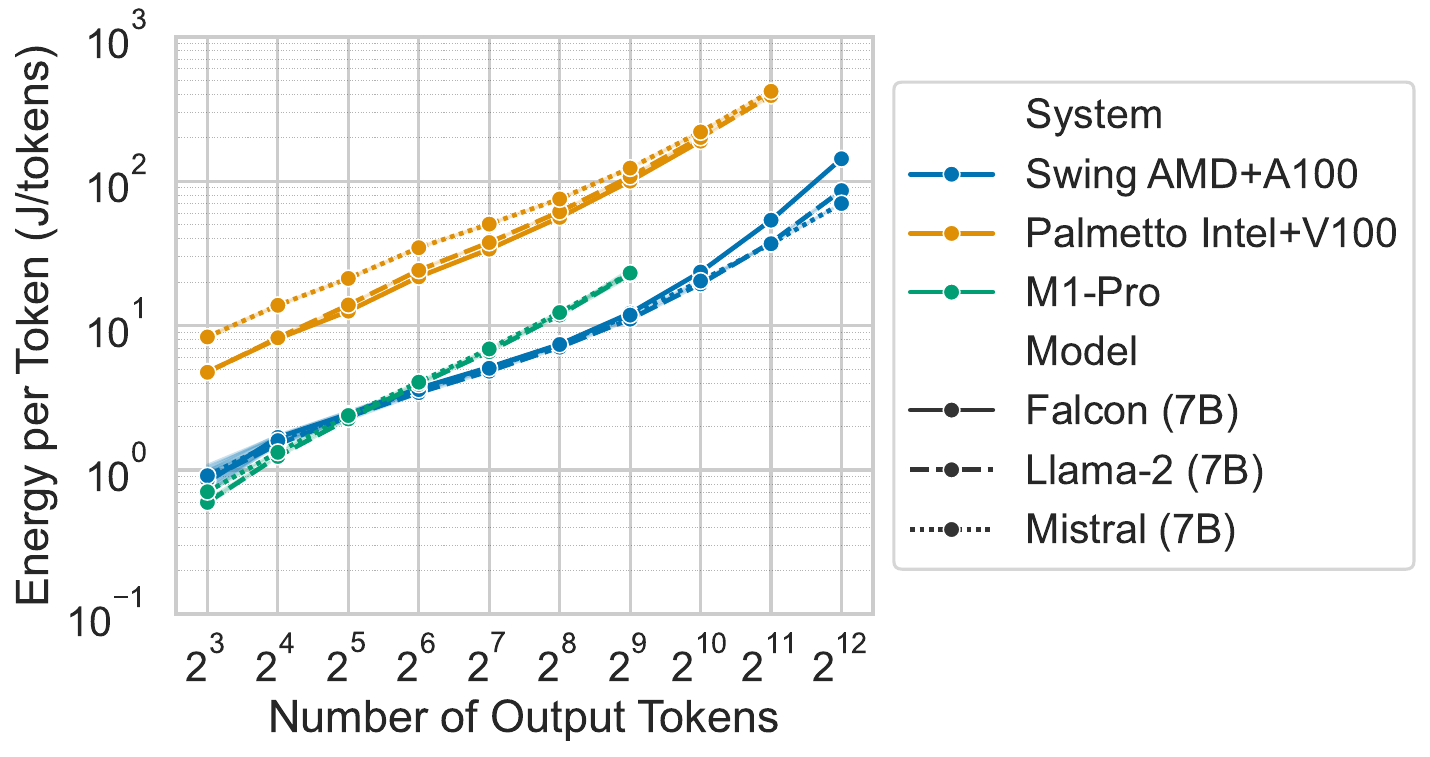}}}
\caption{Performance of Various Systems and Models for Processing Variable Output Tokens--Missing data points in M1-Pro and Palmetto Intel+V100 are due to CUDA out of memory errors. Due to the low variance in the data, error bars are too small to be visible.}
  \label{fig:output-token-results}
  \vspace{-3mm}
\end{figure*}

Runtime significantly increases with the number of output tokens across all systems. As illustrated in Figure~\ref{fig:output-runtime}, the escalation in runtime is pronounced, particularly as the output token count reaches higher magnitudes. This increase is indicative of the substantial computational effort required by LLMs to generate successive tokens.

In Figure~\ref{fig:output-throughput}, we observe a decrease in throughput across all systems as the number of output tokens increases. This trend highlights the inherent computational complexity involved in generating larger sequences of tokens in LLM tasks. As the output token count grows, the system must process each additional token, recalculating the context and updating internal model states~\cite{attention}. This not only increases the total computation per query but also leads to a greater accumulation of processing time per token, which consequently lowers the overall throughput.

Energy consumption per token also shows an increasing trend as the number of output tokens grows. Displayed in Figure~\ref{fig:output-energy-per-token}, this trend underscores the energy-intensive nature of producing larger outputs. Systems such as the M1-Pro, while generally more energy-efficient, begin to consume more energy per token as output demands increase, reflecting the intensive processing involved in output generation.

\subsection{Comparing the Input and Output Analyses}
When comparing Figure~\ref{fig:input-runtime} and Figure~\ref{fig:output-runtime}, we observe that increases in the number of output tokens result in a more considerable increase in runtime than increases in input tokens. The computational complexity of processing input tokens primarily involves encoding the input context, which occurs once per input sequence and follows a more linear computational trajectory. In contrast, generating output tokens is inherently more complex and iterative. Each new output token requires the model to run through all its layers to predict the next token based on an ever-expanding context, which includes both the initial input and all previously generated tokens~\cite{attention}. This ongoing computation involves recalculating attention across an increasing number of tokens, updating hidden states, and generating a probability distribution over the vocabulary for each new token. Consequently, as the number of output tokens grows, the computational load increases significantly, leading to more significant runtime increases than processing input tokens.

The impacts on runtime also translate to the throughput, depicted in Figure~\ref{fig:input-throughput} and Figure~\ref{fig:output-throughput}. There is a noticeable decline in throughput as output tokens increase, more so than input tokens. The decrease in throughput for output tokens is primarily due to the heightened computational requirements for generating subsequent tokens, where each token’s generation slows down as the sequence lengthens. Furthermore, the energy per token also increases as output tokens grow, as shown in our analysis. The energy required to generate each output token becomes significant due to longer passes through the transformer network. We contrast this with the energy consumption when processing input tokens, which, despite increasing, does so at a less steep rate.

\section{Energy-Optimal Hybrid Datacenter for LLM Inference}\label{sec:analysis}
Considering the performance results we collect from LLM inference across multiple systems, we notice that there is an energy-optimal way to construct a hybrid datacenter with a combination of M1 Pro's and A100s. The intuition behind this is that the energy expended per token for the M1 Pro is lower than that of the A100 up to a certain point in the number of input and output tokens as seen in Figures~\ref{fig:input-energy-per-token} and \ref{fig:output-energy-per-token}. However, the energy efficiency characteristics are different when varying the number of input and output tokens, and therefore, we will proceed with separate analyses.

\subsection{Number of Input Tokens Analysis}
Suppose we have a hybrid data center with M1-Pros and A100s. Then, we have some workload for an LLM, a set of queries with some outputs. In such a configuration, we implement a scheduling heuristic based on a cutoff threshold, $T_{in},$ for input token length. This heuristic dictates that queries with $n \leq T_{in}$ tokens are processed on M1 Pro systems, which we have shown have good energy efficiency with handling smaller computational loads. Conversely, queries with $n>T_{in}$ tokens leverage the greater computational ability of A100 GPUs, which offer greater energy-per-token advantages for larger tasks despite their higher power usage. We point out that this is the same method mentioned in the problem formulation in Eqn. \ref{eqn:optimization}, where our queries $Q$ are partitioned into $Q_{M1}$ and $Q_{A100}$ strictly on input and output size.

\begin{figure}[!htb]
  \centering  
\subfigure[{Input Tokens}]
{
\label{fig:token-input-histogram}
\raisebox{-1cm}{\includegraphics[scale=0.25]{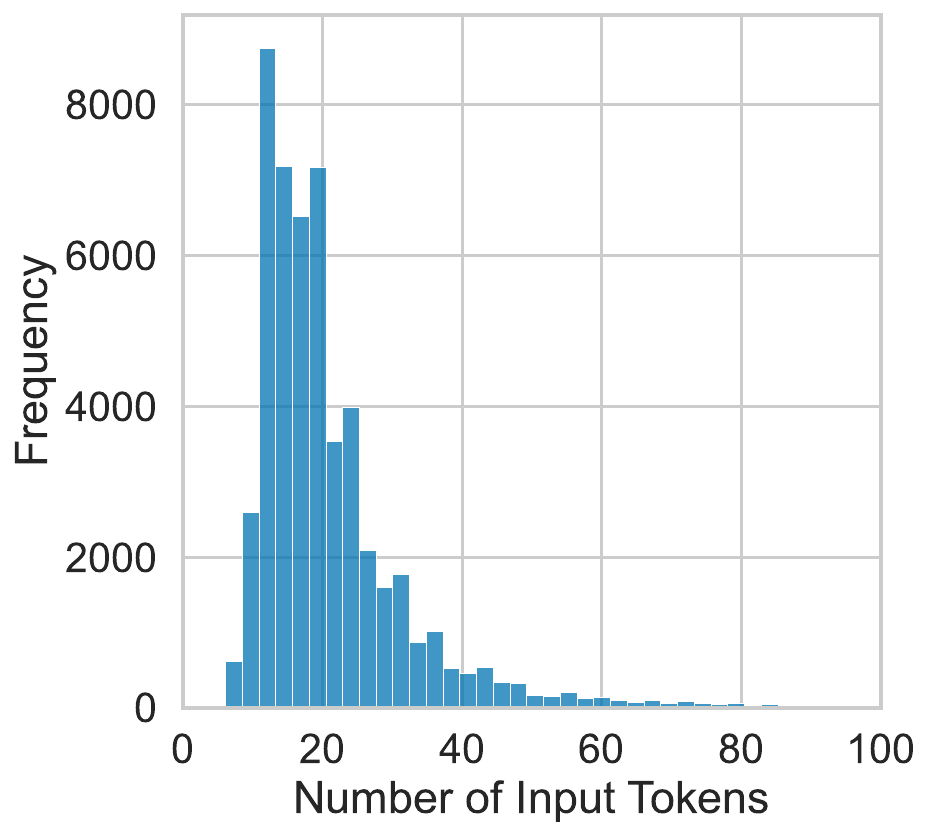}}
}  
\subfigure[{Output Tokens}]
{
\label{fig:token-output-histogram}
\raisebox{-1cm}{\includegraphics[scale=0.25]{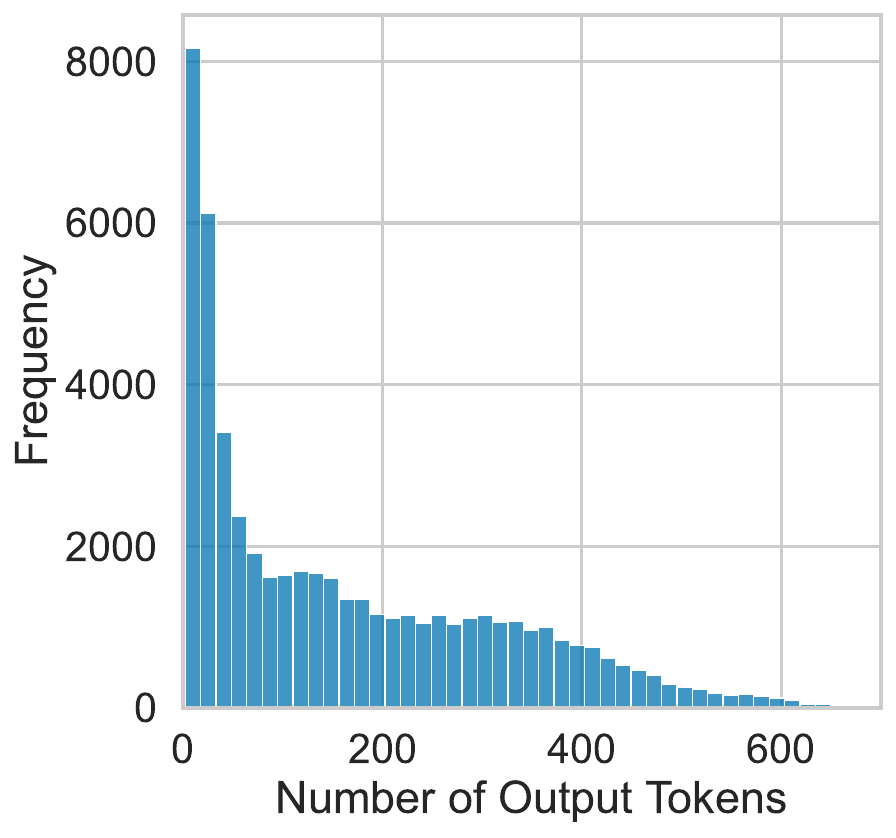}}
}  
\caption{Distribution of Token Counts for Alpaca~\cite{alpaca}}
  \label{fig:distribution-alpaca}
  \vspace{-3mm}
\end{figure}

To find an optimal threshold $T_{in}$ empirically, we analyze the token distribution in prompts from the Alpaca~\cite{alpaca} dataset, a benchmark dataset frequently used in model fine-tuning. This dataset comprises 52K prompts, offering a diverse range of lengths akin to a typical workload in systems like GPT-4~\cite{openai2023gpt4}. The distribution of input tokens, visualized in our analysis (see Fig.~\ref{fig:token-input-histogram}), serves as a proxy for understanding the variegated nature of LLM workloads.

\begin{figure}[!htb]
  \centering  
  \captionsetup{justification=centering}
\subfigure[{Energy Consumption for Changing $T_{in}$\hspace{-1.1cm}}]
{
\label{fig:input-energy-hybrid}
\raisebox{-1cm}{\includegraphics[scale=0.34]{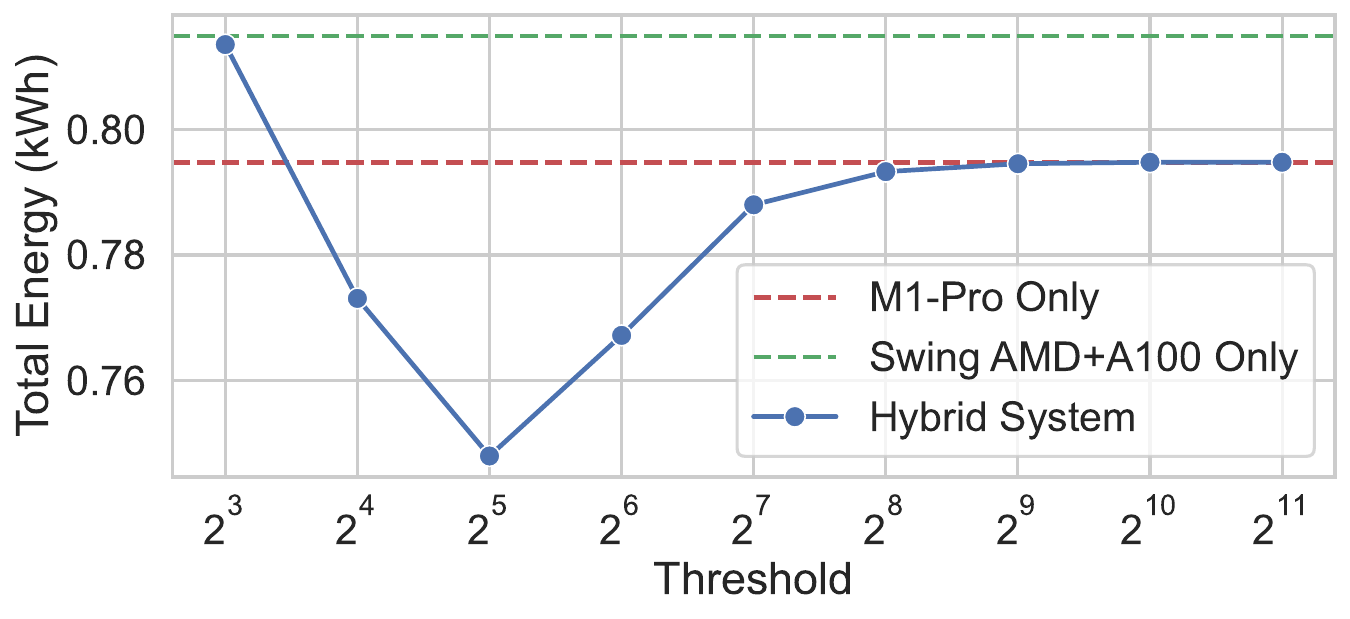}}
}  
\subfigure[{Runtime for Changing $T_{in}$\hspace{-1cm}}]
{
\label{fig:input-runtime-hybrid}
\raisebox{-1cm}{\includegraphics[scale=0.34]{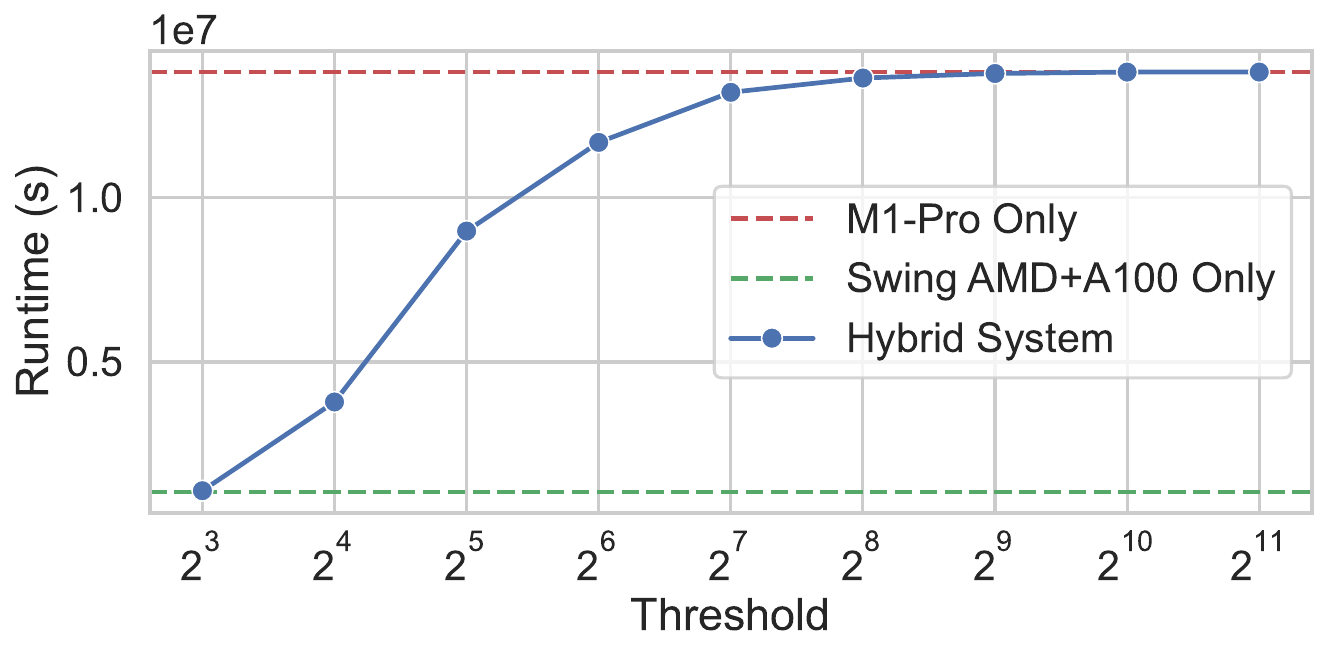}}
}  
\caption{Performance of Hybrid Datacenter for Input Tokens Processing Alpaca--Dashed line shows the value for using only one kind of hardware for inference}
  \label{fig:input-hybrid-datacenter}
  \vspace{-3mm}
\end{figure}

The energy component of our cost function, split over the token threshold, is as follows:
\begin{equation}\label{eqn:performance-model-input}
    E_{Total, in} = \sum_{m = 1}^{T_{in}}mf_{in}(m)E_{M1, in}(m) + \sum_{m = T_{in}+1}^M mf_{in}(m)E_{A100, in}(m),
\end{equation}
where $E_{Total, in}$ represents the total energy consumption for a given dataset of input lengths $m$ with corresponding frequencies $f_in(m),$ and $E_{M1, in}(m)$ and $E_{A100, in}(m)$ denote the mean energy per token for varying the input token size for the M1-Pro and A100 systems, respectively. Utilizing this model with our dataset enables the approximation of total energy consumption for various threshold settings, offering insights into the energy dynamics of hybrid datacenter operation. In Figure~\ref{fig:input-hybrid-datacenter}, we show the energy and runtime simulation results of performing inference for the input token sizes from the Alpaca dataset.

Our findings indicate that a threshold of 32 tokens strikes an optimal balance, significantly reducing energy consumption by relegating the inference of shorter queries to the more energy-efficient M1 Pro systems. This policy not only capitalizes on the inherent energy efficiency of the M1 Pro for smaller tasks but also reserves the computational might of the A100 for queries that necessitate its robust capabilities. However, it's important to note that this energy optimization comes at the cost of increased runtime. 

\subsection{Number of Output Tokens Analysis}
We want to use the same scheduling heuristic and performance model to determine a threshold $T_{out}$ for the number of output tokens. Except this time, we have different frequencies $f_{out}(n)$ for the $n$ output tokens and different mean energy per token for varying the output token size, $E_{M1, out}(n)$ and $E_{A100, out}(n).$ We also utilize the distribution of the number of output tokens in the Alpaca dataset (see Fig.~\ref{fig:token-output-histogram}). We revise our performance model as follows:

\begin{equation}\label{eqn:performance-model-output}
\begin{split}
    E_{Total, out} = \sum_{n = 1}^{T_{out}} nf_{out}(n)&E_{M1, out}(n)\\
    &+ \sum_{n = T_{out}+1}^N nf_{out}(n)E_{A100, out}(n).
\end{split}
\end{equation}

\begin{figure}[!htb]
  \centering  
  \captionsetup{justification=centering}
\subfigure[{Energy Consumption for Changing $T_{out}$\hspace{-1cm}}]
{
\label{fig:output-energy-hybrid}
\raisebox{-1cm}{\includegraphics[scale=0.34]{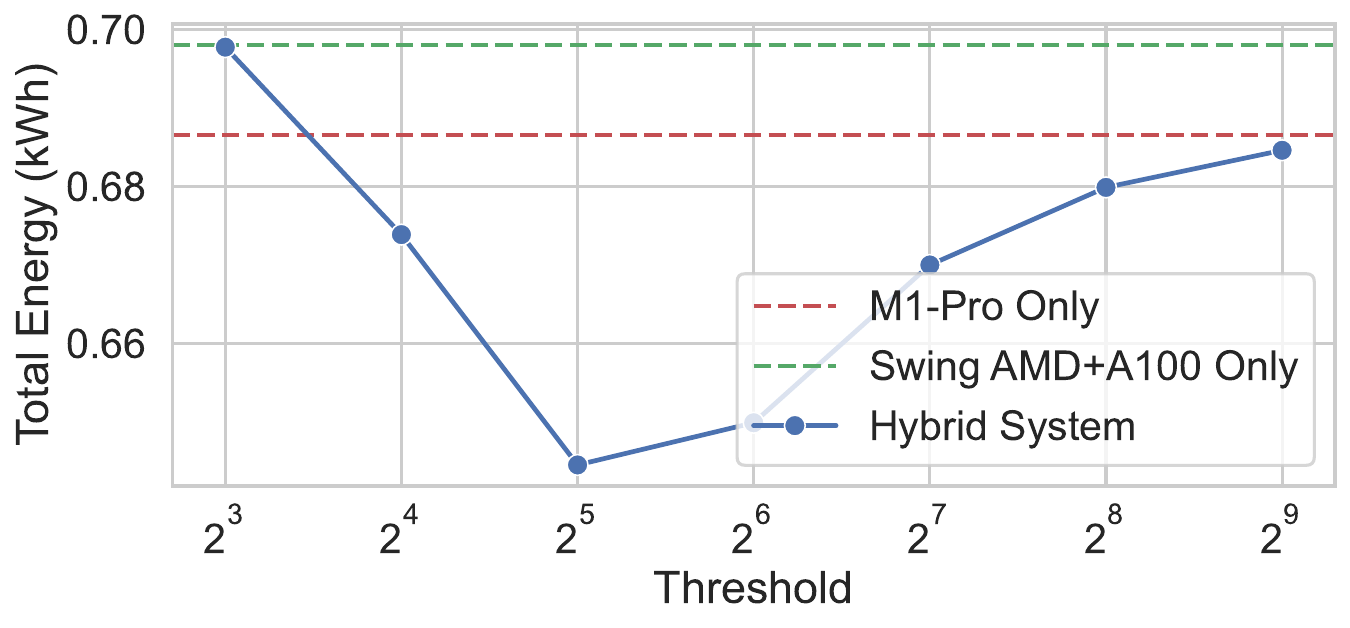}}
}  
\subfigure[{Runtime for Changing $T_{out}$\hspace{-1cm}}]
{
\label{fig:output-runtime-hybrid}
\raisebox{-1cm}{\includegraphics[scale=0.34]{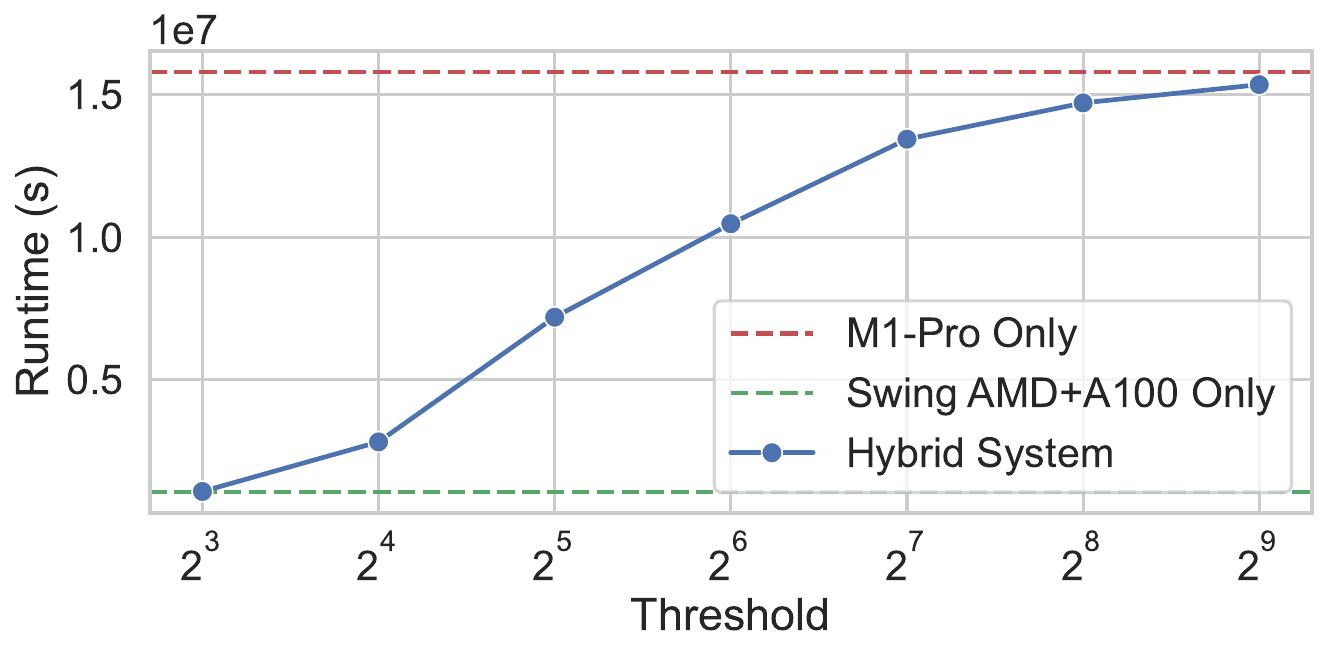}}
}  
\caption{Performance of Hybrid Datacenter for Output Tokens Processing Alpaca -- Dashed line shows the value for using only one kind of hardware for inference}
  \label{fig:output-hybrid-datacenter}
  \vspace{-3mm}
\end{figure}

As the M1 Pro could only generate up to 512 tokens of a response, we only test $T_{out}$ up until this point. 
In Figure~\ref{fig:output-hybrid-datacenter}, we show the energy and runtime simulation results of performing inference for the input token sizes from the Alpaca dataset.

Fig.~\ref{fig:output-runtime-hybrid} and Fig.~\ref{fig:output-energy-per-token} assess the energy consumption and runtime implications of various threshold settings for output generation. Our findings suggest that although higher thresholds may leverage the M1 Pro's energy efficiency for smaller outputs, there is an optimal point at 32 output tokens that minimizes energy consumption.

\subsection{Balancing Energy Efficiency and Runtime Performance}
Our analysis of both input and output token processing within a hybrid, heterogeneous datacenter framework has led to the identification that with certain thresholds at \(T_{input} = 32\) and \(T_{output} = 32\),
we can strategically allocate tasks to M1 Pro systems or A100 GPUs based on token count, optimizing for energy efficiency.

Shifting the token distribution leverages the M1 Pro's superior energy efficiency for input and output tasks up to the threshold, beyond which we utilize the A100's computational power. This policy saves energy as smaller-token tasks are handled by the more efficient M1 Pro for outputs up to the threshold.
However, this energy optimization comes at the expense of increased runtime, which is particularly noticeable in output token generation where the M1 Pro, despite its efficiency, does not match the A100's speed.

The energy-runtime trade-off presents a favorable scenario for applications that have low runtime sensitivity. For instance, batch processing of LLM tasks, such as overnight data analyses or non-time-critical computations, can benefit significantly from this energy-efficient configuration. Similarly, free or not directly monetized services, where the cost of computation impacts operational sustainability, stand to gain from minimizing energy expenditures even at the cost of longer processing times.

This approach also opens discussions on Quality of Service (QoS) for LLMs, an area that still needs to be explored~\cite{wang2024efficient,databricks}. Traditional QoS metrics often prioritize speed and reliability, but energy efficiency may also become a critical QoS dimension for LLM applications, particularly in energy-constrained or cost-sensitive scenarios.

\section{Related Work}\label{sec:related-works}

\subsection{Hybrid and Energy Efficient Heterogeneous Data Centers}
Recent studies in optimizing data center architectures for deep learning have highlighted the necessity of energy-efficient scheduling and task allocation across diverse hardware. Gu et al.~\cite{gu2023powerflow} explore GPU clusters' energy-efficient scheduling, revealing substantial improvements in power utilization without considering diverse GPU types for different task requirements. This work highlights a gap in understanding how various GPU configurations could enhance energy efficiency further. Similarly, Patel et al.~\cite{patel2023hybrid} demonstrate the benefits of hybrid computing environments, emphasizing FPGA over GPU diversity. This focus leaves room to explore the specific impacts of different GPU classes in such settings.

In the realm of LLMs, Zhao et al.~\cite{zhao2024llm} introduce strategies like phase-aware partitioning and adaptive quantization in heterogeneous clusters but do not integrate energy considerations into their analysis, which is crucial for understanding the real-world applicability of these models in power-sensitive environments. On the other hand, Radovanović et al.~\cite{radovanovic2022carbon} and Chien et al.~\cite{chien2023reducing} discuss broader aspects of carbon-aware computing and reducing the carbon impact of AI inference, respectively. These works emphasize the importance of node/device-level energy metrics, often overlooked in typical LLM deployment strategies, thus underscoring the need for detailed energy consumption profiling across different models and hardware types.

\subsection{LLM Inference as a Service}
Further focusing on energy consumption, Hu et al.~\cite{helios2021hu} analyze deep learning workloads in GPU datacenters, offering insights into energy conservation strategies through workload scheduling. This research aligns with our objectives by confirming the critical role of scheduling in reducing energy footprints. Anderson et al.~\cite{anderson2023treehouse} propose carbon-aware datacenter software that could complement physical hardware adjustments by making energy and carbon metrics visible to application developers, encouraging more energy-efficient coding practices.

Addressing service quality, Wang et al.~\cite{wang2024efficient} study the efficiency and reliability of LLM serving, highlighting the challenges of maintaining high-quality service while managing computational loads effectively. This perspective is pertinent as it underscores the tradeoff between performance and energy efficiency, which is central to our study. Lastly, Desislavov et al.~\cite{radosvet2023trends} provide a timely examination of trends in AI inference energy consumption, arguing that while performance has increased dramatically, energy consumption has not escalated at the same pace, thanks to hardware optimizations and algorithmic innovations. This outlook is necessary as it suggests the potential for further optimizations in LLM inference tasks, which are typically energy-intensive.

\section{Conclusions}\label{sec:conclusions}
By carefully analyzing the energy and runtime of heterogeneous compute hardware to host LLMs,
we show that a hybrid, heterogeneous datacenter and a cost-based scheduling framework
can allocate LLM tasks to accelerators that are best suited to run them, in terms of
energy efficiency and computational performance. This
decision is based simply on the size of input and output tokens, making the decision process easy to 
integrate into existing workloads. Our cost function allows us to balance energy consumption against runtime , providing a method to quantitatively assess and manage trade-offs in real-time.
By demonstrating that data-driven approaches can mitigate the energy impact of serving LLM inference, we hope to pave the way for more energy-efficient and environmentally friendly technology infrastructures.

\section*{Acknowledgment}
We gratefully acknowledge the computing resources provided on Swing and Palmetto, both high-performance computing clusters operated by the Laboratory Computing Resource Center at Argonne National Laboratory and Clemson University, respectively. During this work GW was supported by a Churchill Scholarship.

\bibliographystyle{acm}
\bibliography{e2dc}

\end{document}